# 4D visualization of the photoexcited coherent magnon by an X-ray free electron laser

Hoyoung Jang[1,2], Hiroki Ueda[3], Hyeong-Do Kim[1], Minseok Kim[1], Kwang Woo Shin[4], Kee Hoon Kim[4], Sang-Youn Park[1], Hee Jun Shin[1], Pavel Borisov[5,6], Matthew J. Rosseinsky[6], Dogeun Jang[1], Hyeongi Choi[1], Intae Eom[1,2], Urs Staub[3] and Sae Hwan Chun[1,2]*

[1]*Pohang Accelerator Laboratory, Pohang, Gyeongbuk 37673, Republic of Korea*
[2]*Photon Science Center, POSTECH, Pohang, Gyeongbuk 37673, Republic of Korea*
[3]*Swiss Light Source, Paul Scherrer Institute, 5232 Villigen-PSI, Switzerland*
[4]*Center for Novel States of Complex Materials Research, Department of Physics and Astronomy, Seoul National University, Seoul 08826, Republic of Korea*
[5]*Department of Physics, Loughborough University, Loughborough, LE11 3TU, United Kingdom*
[6]*Department of Chemistry, University of Liverpool, Liverpool, L7 3NY, United Kingdom*
* Corresponding author: pokchun81@postech.ac.kr

**X-ray free electron lasers (XFEL) create femtosecond X-ray pulses with high brightness and high longitudinal coherence allowing to extend X-ray spectroscopy and scattering techniques into the ultrafast time-domain. These X-rays are a powerful probe for studying coherent quasiparticle excitations in condensed matter triggered by an impulsive optical laser pump. However, unlike coherent phonons, other quasiparticles have been rarely observed due to small signal changes and lack of standards for the identification. Here, we exploit resonant magnetic X-ray diffraction using an XFEL to visualize a photoexcited coherent magnon in space and time. Large intensity oscillations in antiferromagnetic and ferromagnetic Bragg reflections from precessing moment are observed in a multiferroic *Y*-type hexaferrite. The precession trajectory reveals that a large, long-lived, photoinduced magnetic-field changes the net magnetization substantially through the large-amplitude of the magnon. This work demonstrates an efficient XFEL probe for the coherent magnon in the spotlight for opto-spintronics application.**

The quantum mechanical wave-particle duality is manifested in condensed matter as quasiparticles corresponding to collective modulations of physical degrees of freedom. A widely known example is phonons formed by collective vibrations of atoms. Femtosecond optical lasers enable time-resolved optical measurements to probe the coherent lattice vibration excited by the optical laser pump[1,2]. Femtosecond X-ray sources further advance the investigation of coherent phonons, allowing X-ray diffraction to directly quantify the atomic-scale motions in the time domain[3-7]. Direct access to interatomic potential energy surface becomes available by this experimental technique, which elucidates non-equilibrium processes forming transient material phases[5,6,8].

Magnons are quasiparticles in the magnetic system, i.e., spin wave excitations representing a collective precession of magnetic moments in an ordered magnetic structure. In particular, the magnons in the antiferromagnet have gained a growing interest from magnonics research[9,10] as their high precession frequency and capability to code/decode information in the precessing direction, amplitude, and phase are considered key attributes for high-speed information carriers. The antiferromagnet itself has also drawn attention from spintronics research[11,12] due to great stability against magnetic-field perturbation and sizable magnetotransport effect suited for magnetic memories. An emerging research field of opto-spintronics[13-15] targets ultrafast manipulation of the magnetic moment through the photoexcited coherent magnon. For the antiferromagnet, however, the zero net-magnetization resulting from the compensated sublattice moments permits only limited access to the precessing moment by magneto-optical techniques[14,16,17].

Time-resolved resonant magnetic X-ray diffraction directly probes the precessing magnetic moment manifested as an oscillation in the magnetic Bragg peak intensity[18-20]. The XFEL is advantageous for this type of experiment that demands high temporal resolution and intense X-ray flux[21]. However, the XFEL experiment has struggled to measure small oscillation amplitude of the coherent magnon[18] and set the criterion to experimentally discriminate the relevant features from other effects such as coherent phonons, demagnetization, and melting magnetic-order[22-26]. In this work, we designed and carried out an experiment to demonstrate the versatile capability of the XFEL probe for the photoinduced coherent magnon. The resonant magnetic X-ray diffraction process offers distinct X-ray polarization channels that view conjugate components of the precessing moment. The oscillation profiles varied upon the moment directions with respect to the polarization channels establish the criterion to discern the coherent magnon from the other aforementioned effects. We also exploit the fact that the X-ray absorption length dramatically changes near the absorption edge, to maximize the contrast of the intensity oscillation by matching the penetration depths between the optical laser pump and X-ray probe[27]. In terms of the model material system, a *Y*-type hexaferrite is chosen because its canted *AFM* order[28,29] is a known platform displaying large-amplitude coherent magnons[30-32]. The substantial magnetic-scattering cross-section of the $Fe^{3+}$ ion near Fe $L_3$ absorption edge (~710 eV) leads the magnetic Bragg reflection intensities to be comparable to the structural Bragg reflections[33], which in turn facilitates capturing the intensity oscillation from the enhanced signal. All these tactics combined, we succeed to visualize a coherent *AFM* magnon in four dimensions (4D), i.e., three-dimensional space and time. This work establishes a unique methodology to study pathways of ultrafast non-equilibrium processes in photoexcited magnets.

**Photoinduced intensity oscillations in magnetic Bragg reflections**

The *Y*-type hexaferrite studied in this work is $Ba_{0.5}Sr_{1.5}Zn_2(Fe_{1-x}Al_x)_{12}O_{22}$ ($x = 0.08$) (BSZFAO) that exhibits a transverse conical magnetic structure under magnetic field in the basal plane below ~260 K (Fig. 1a)[29]. This canted *AFM* structure consists of *FM* (ferromagnetic) and *AFM* components, from net magnetic moments $\mu_L$ and $\mu_S$ of the structural *L* (large) and *S* (small) blocks, respectively. When the magnetic field is applied along the *x* axis in the *ab* plane, the *FM* component of $\mu_L$ (***m*** = $\mu_{L1}$ + $\mu_{L2}$) aligns along the field and the *AFM* component of $\mu_L$ (***l*** = $\mu_{L1}$ - $\mu_{L2}$) along the orthogonal *y* axis in the *ab* plane. Meanwhile, $\mu_S$ directs the *FM* component opposite to the field and the *AFM* component along the *z* axis (|| [001]). The individual sublattices of $\mu_L$ and $\mu_S$ contribute to *FM* and *AFM* Bragg intensities of (0 0 3*n*) and (0 0 3*n*±1.5) reflections with integer *n*, respectively, accessible due to the long *c* axis parameter (*c* = 43.3 Å, *a* = *b* = 5.85 Å) by resonant soft x-ray diffraction near the Fe $L_3$ edge ~710 eV (Figs. 1b & c). The horizontal scattering geometry with the magnetic field applied normal to the scattering plane (Fig. 1d) leads the π-π' and π-σ' polarization channels to view the structure factors contributed from the *FM* and *AFM* components projected in the *x* axis and incident X-ray direction ($k_i$), respectively.

The optical laser pump and X-ray probe experiment is employed to investigate temporal changes of the magnetic Bragg reflections. The optical laser (wavelength = 400 nm) excites the valence electrons across the band gap (< 3.1 eV)[34,35], while the X-ray at $E_i$ = 702 eV, with an energy of a few eV below the Fe $L_3$ edge, penetrates deep into the material (penetration depth = 363 nm) and probe the bulk properties (see Methods). Figures 2a-c present *FM* (0 0 3) and *AFM* (0 0 1.5) & (0 0 4.5) Bragg reflections as a function of time delay after the photoexcitation. It is evident that these Bragg peaks display multiple oscillation cycles with a period of 24±0.5 ps. Their Fourier transformation indicates a single oscillation frequency centered at 41.7±0.9 GHz (≈ 0.17 meV), implying a quasiparticle with the corresponding energy.

Quasiparticles such as coherent phonons and magnons are able to oscillate Bragg reflection intensities. We first examine the coherent phonon as a possible origin for the oscillations. The resonant *FM* (0 0 3) Bragg peak contains structure factors not only from the *FM* component of the canted *AFM* structure, but also from the crystal structure and the orbital asphericity. The coherent phonon oscillation would be present even for a non-resonant condition suppressing the magnetic contribution, but it turns out that no oscillation is found in the transient intensity (see Supplementary Fig. 3). Our attention is turned to the magnetic origin by noticing that reversing magnetic field inverts the oscillation profiles in the *AFM* (0 0 4.5) Bragg reflection as shown in Fig. 2d. Inspection of possible precession motions reveals that an *AFM* magnon precession, in which $\mu_{L1}$ and $\mu_{L2}$ precess counter-clockwise around the precession axes tilted further away from the *x* axis (Fig. 2f), explains the oscillation profiles for the *AFM* (0 0 4.5) and *FM* (0 0 3) Bragg intensities (see Supplementary Figs. 7-10). Figure 2g illustrates the *AFM* precession and its projection on the scattering plane for the case of $H_{ext}$ ∥ +*x*. It is noticed that the *AFM* (0 0 4.5) intensity ∝ $|l \cdot k_i|^2$ minimizes/maximizes at every first/third quarter period, while the *FM* (0 0 3) intensity ∝ $|m \cdot x|^2$ minimizes/maximizes at every half/full period. Thus, the *FM* and A*FM* profiles exhibit cosinusoidal and inverted sinusoidal forms, respectively. [The slight phase difference is neglected in the discussion.] This coherent *AFM* magnon precession is corroborated by reversing $H_{ext}$ ∥ -*x* (Fig. 2h). The $\mu_L$ and its precession trajectory would rotate by 180˚ around the *z* axis, and thereby the *AFM* and *FM* profiles are expected to be inverted and maintained, respectively. These behaviours are indeed confirmed by the observation shown in Figs. 2d & e.

**An effective magnetic Hamiltonian model calculation**

A magnetic Hamiltonian model provides further understanding of the coherent *AFM* magnon. This Hamiltonian is built from the *L* & *S* blocks[28,35,36]:

$$\mathcal{H} = J_{LS} \sum_{i,j=1,2} \vec{\mu}_{Si} \cdot \vec{\mu}_{Lj} + 2J_{LL}\vec{\mu}_{L1} \cdot \vec{\mu}_{L2} + 2J_{SS}\vec{\mu}_{S1} \cdot \vec{\mu}_{S2}$$
$$+ D_L \sum_{i,j=1,2}(\mu_{Li}^z)^2 + D_S \sum_{i,j=1,2}(\mu_{Si}^z)^2 - H_x \sum_{i,j=1,2}(\mu_{Li}^x + \mu_{Si}^x) \quad (Eq.\ 1)$$

with $J_{LS}$, $J_{LL}$, and $J_{SS}$ being exchange interaction constants of *L* and *S*, *L* and *L*, and *S* and *S* blocks, respectively; $D_L$ and $D_S$ are the magnetic anisotropy constants of *L* and *S* blocks along the *z* axis (|| [001]), respectively, and $H_x$ is the sum of internal and external magnetic fields along the *x* axis. We calculate the static magnetization ($M_x$) vs. magnetic field (|| +*x*) and obtain most of the relevant parameters, except $D_L$ and $D_S$, in the Hamiltonian by matching the experimental data shown in Fig. 3a (see Methods). The magnetic anisotropy constants can be determined from the dynamics of magnetic moments, e.g., magnon energies. We solve the Landau-Lifshitz-Gilbert equation and find two solutions corresponding to *AFM* normal modes that embody distinct *AFM* precessions of $\mu_{L1}$ and $\mu_{L2}$, and $\mu_{S1}$ and $\mu_{S2}$ (see Methods and Supplementary Fig. 2). The coherent magnon with the frequency of 41.7 GHz (≈ 0.17 meV) is assigned to the lower energy *AFM* magnon[35] by noting that a number of *Y*-type hexaferrites have another *AFM* magnon mode in the higher frequency range of ~0.5 - 1.5 THz[37,38].

Figure 3d illustrates relative magnitudes of $\mu_{L1}$ and $\mu_{S1}$ precessions for the *AFM* magnon calculated from the magnetic Hamiltonian. This magnon has dominant $\mu_L$ precession, of which trajectory is nearly isotropic in the surface normal to the precession axis. On the other hand, the $\mu_S$ precession is insignificant while elongated along the *x* axis. This calculation result verifies our data interpretation with the $\mu_L$ precession.

**Quantification of the precessing magnetic moment**

A unique capability of the time-resolved resonant magnetic X-ray diffraction is a direct quantification of the precessing magnetic moment. Figure 3b shows the *AFM* (0 0 4.5) intensity at $E_i$ = 711 eV (for matching the X-ray penetration depth with that of the laser ~39 nm) which exhibits a 28 % initial oscillation amplitude of the static value at the laser fluence = 1.0 mJ/cm². This change corresponds to the situation that $\mu_{L1}$ ($\mu_{L2}$) moves out of the *xy* plane by 4° towards +*z* (-*z*) axis at the first quarter of the oscillation period, respectively. The estimation from the *FM* (0 0 3) intensity with the initial oscillation amplitude of 18 % (Fig. 3c) implies that when $\mu_{L1}$ and $\mu_{L2}$ come back on the *xy* plane at the half period, they tilt away from each individual precession axis by 8° towards -*x* axis. We note that the net magnetization $M_x$ changes substantially during this precession. From the $M_x$ vs. $H_{ext}$ relation derived from the magnetic Hamiltonian, it is deduced that the initial $\mu_{L1}$ and $\mu_{L2}$ point away from the *x* axis by 60° (Figs. 3a & e). Thus, the precession results in $\Delta M_x$ = -1.8 $\mu_B$/f.u. (~26.4 % of the static $M_x$) at the first half period and $\Delta M_x$ = -0.9 $\mu_B$/f.u. at longer time delays when the magnetic system reaches a transient quasi-equilibrium (Fig. 3e). Such unusual large changes in the magnetization suggest a paradigm shift from the view that optically excited coherent magnon magnitudes are too small to change the macroscopic magnetization substantially.

**Discussion**

The precession trajectory distinguishes the mechanism generating the coherent magnon. A linearly polarized laser is able to promote a coherent magnon through either ICME (inverse Cotton-Mouton effect)[31,32] or DECM (displacive excitation of coherent magnon)[39] processes; the former involves an impulsive magnetic torque that rotates the magnetic moment around the original equilibrium direction, while the latter forms a new quasi-equilibrium direction around which the moment precesses (as analogous to the displacive excitation of coherent phonon[2]). The ICME and DECM are differently affected by the amount of optical absorption, and are favored in optically transparent and opaque magnetic materials, respectively[31]. The *AFM* magnon trajectory identified in Fig. 3e indicates the presence of a new quasi-equilibrium direction, and thereby is consistent with the DECM process. The substantial optical absorption from the above-band-gap excitation supports the DECM for the generation of the coherent magnon. [Thermal origin for the coherent magnon is ruled out as the initiation of the magnon precession is almost instantaneous, and faster than the time scale of lattice thermalization ~10 ps (see Supplementary Fig. 4).]

The *AFM* magnon precession trajectory indicates that the photoexcitation acts to effectively reduce the magnetic field. It follows that an effective photoinduced field ($H_{photo}$ || -x) can be identified by matching the quasi-equilibrium position of the magnetic moment to the static one at lower $H_{ext}$. Figure 4c summarizes the $H_{photo}$ dependence on the laser fluence. For low fluences < 1.0 mJ/cm$^2$, the $H_{photo}$ follows a linear line with a slope ~2 kOe.cm$^2$/mJ, consistent with the proportionality known for the DECM process[39]. Above 1.0 mJ/cm$^2$, it becomes hard to unambiguously determine the quasi-equilibrium moment positions and also the $H_{photo}$, because the intensity oscillation profile is distorted by other effects such as ultrafast melting of the *AFM* order within 1 ps, red-shifting of the magnon frequency due to a substantial increase of temperature after 10 ps, and the irregular oscillation amplitude possibly due to enhanced magnon-magnon or magnon-phonon scatterings (Fig. 4a). Despite these complications, the observation that initial magnon oscillation amplitude for 2.6 mJ/cm$^2$ is close to the one for 1.0 mJ/cm$^2$ implies the similar strength of the $H_{photo}$. We note that the large, long-lived $H_{photo}$ reaching ~1.75 kOe accounts for the large change of $M_x$ through the coherent magnon (Fig. 3). This field is as strong as that of permanent rare-earth magnets, evidencing large opto-magnetic coupling in the BSZFAO.

The quantification of the effective photoinduced magnetic field offers a direct access to the magnetic potential energy surface. The magnetic potential energy calculated from the magnetic Hamiltonian model (*Eq*. 1) for the lowered magnetic field effectively describes the transient potential energy surface. Figure 4b presents the potential energies calculated for the effective magnetic field $H_{eff}$ = 2 kOe and 0.25 kOe towards +x axis. The latter corresponds to the transient case for 1.0 mJ/cm$^2$ ($H_{eff}$ = $H_{ext}$ - $H_{photo}$). It is understood that the photoexcitation places the magnetic system on high energy side of the energy surface and starts to move towards the new energy minimum by generating magnons. In Fig. 4b, the initial magnon precession energy is counted as ~16 $\mu_B$·kOe/f.u., which corresponds to ~250 nJ/cm$^2$ for the sample volume mainly absorbing the laser photons. The ratio of this energy to the laser fluence = 1.0 mJ/cm$^2$ gives the photoexcitation efficiency of ~2.5 × 10$^{-4}$ for the BSZFAO. Above this fluence, the thermal effects start to play a role (Fig. 4a). Thus, the magnitude of the photoexcitation efficiency for the laser fluence = 1.0 mJ/cm$^2$ suggests an upper boundary for creating the non-thermal transient state. This identification of energetics of the coherent magnon excitation and transient magnetic state contributes to establishing predictive models to describe non-equilibrium processes in the magnetic system.

**Conclusion**

The presented 4D visualization of the coherent magnon realizes a long-anticipated capability of XFELs for characterizing coherent quasiparticle excitations. We find that the BSZFAO exhibits remarkably large modulation of the *FM* magnetization by the photoexcited *AFM* magnon. The notion that this *Y*-type hexaferrite is a multiferroic with ferroelectricity and ferrimagnetism coupled each other[40-42], suggests application prospects not only for opto-spintronics, but also the optical control of magnetoelectrics[43,44]. This work demonstrates the advantages of resonant magnetic X-ray diffraction brought into the time domain, and further inspires research on novel quasiparticle excitations in materials via XFELs[45,46].

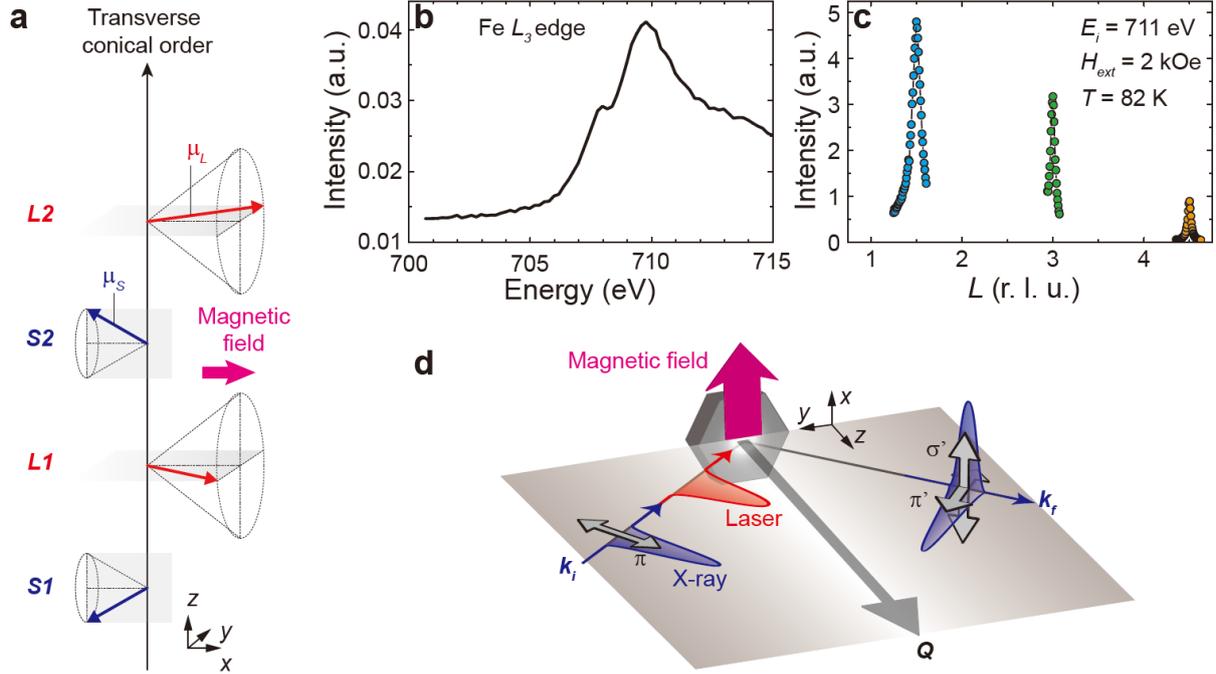

**Fig. 1| Magnetic structure of the *Y*-type hexaferrite and experimental layout for the time-resolved resonant magnetic x-ray diffraction.** (**a**) The canted *AFM* magnetic structure represented by net magnetic moments of the *L* ($\mu_L$, red arrow) and *S* ($\mu_S$, blue arrow) blocks under the external magnetic field, $H_{ext} \parallel +x$. The $\mu_L$ and $\mu_S$ are constrained in horizontal and vertical planes (grey), respectively. Dotted black frames are a guide to the eye for conceiving the transverse conical structure. (**b**) X-ray absorption spectrum at the Fe $L_3$ edge. (**c**) *L*-scan covering *FM* (0 0 3) and *AFM* (0 0 1.5) & (0 0 4.5) Bragg peaks at *T* = 82 K using incident X-ray photon energy, $E_i$ = 711 eV. (**d**) Layout of the time-resolved resonant magnetic X-ray diffraction. The horizontal scattering plane (grey gradient) is defined by incident ($k_i$) and outgoing ($k_f$) wavevectors. Optical laser (red) and X-ray (blue) pulses are collinearly incident on the sample with the polarizations lying in the scattering plane, i.e. p- and π-polarizations, respectively, while the diffracted X-ray pulse has both π'- and σ'-polarizations.

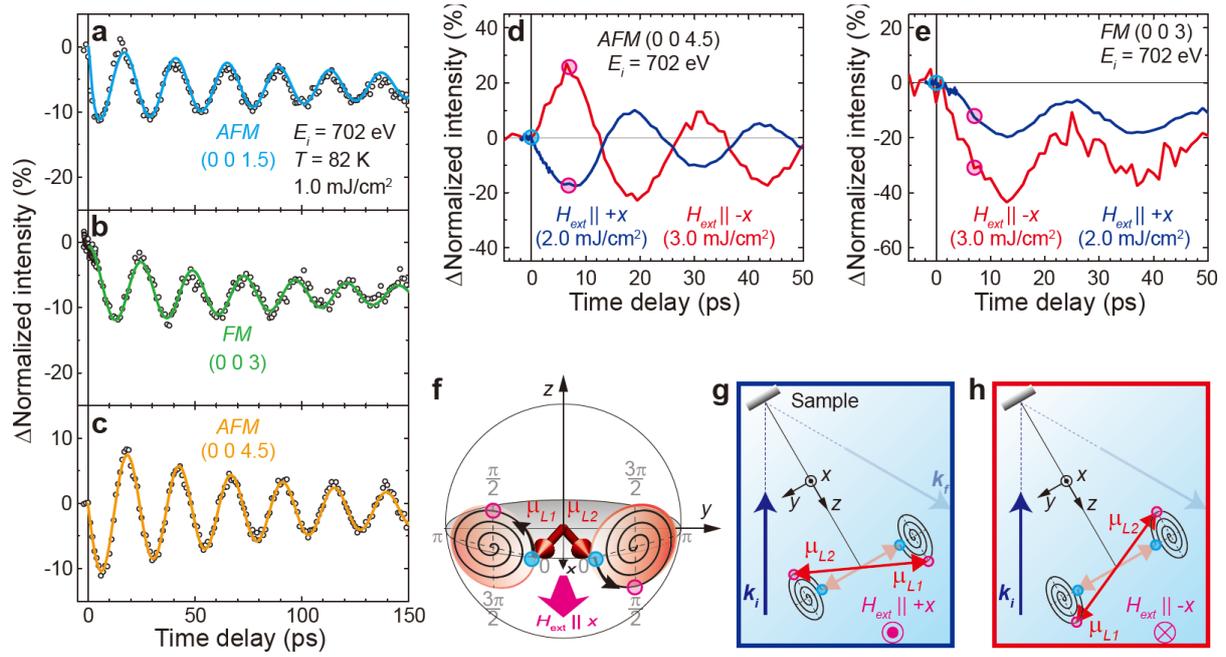

**Fig. 2| Transient magnetic Bragg peak intensities.** (**a**) *AFM* (0 0 1.5), (**b**) *FM* (0 0 3), and (**c**) *AFM* (0 0 4.5) reflections at $T$ = 82 K for the laser fluence = 1.0 mJ/cm$^2$ using $E_i$ = 702 eV. The transient data (symbol) are normalized to the intensity at negative time delay = -1 ps for each reflection. The solid lines are fitting results combining a damped oscillation and an exponential decay (see Supplementary Table 1). (**d**) *AFM* (0 0 4.5) and (**e**) *FM* (0 0 3) reflections for $H_{ext}$ ∥ +$x$ (blue, 2.0 mJ/cm$^2$) and -$x$ (red, 3.0 mJ/cm$^2$) at $T$ = 82 K. The colored circles correspond to the same colored $\mu_L$ directions on the precession trajectory shown in (**f**). (**f**) Schematic illustration of $\mu_{L1}$ and $\mu_{L2}$ that precess counter-clockwise (denoted by black arrows) around the precession axes tilted further away from the $x$ axis. The spiral lines represent the precession trajectories towards the new equilibrium directions. Top view of the experimental geometries for $H_{ext}$ ∥ +$x$ (**g**) and -$x$ (**h**). The $\mu_L$ and its precession trajectory are projected on the horizontal scattering plane. The $\mu_L$ at the beginning of the precession and at the first quarter of the oscillation period are denoted by the light red and dark red arrows, respectively.

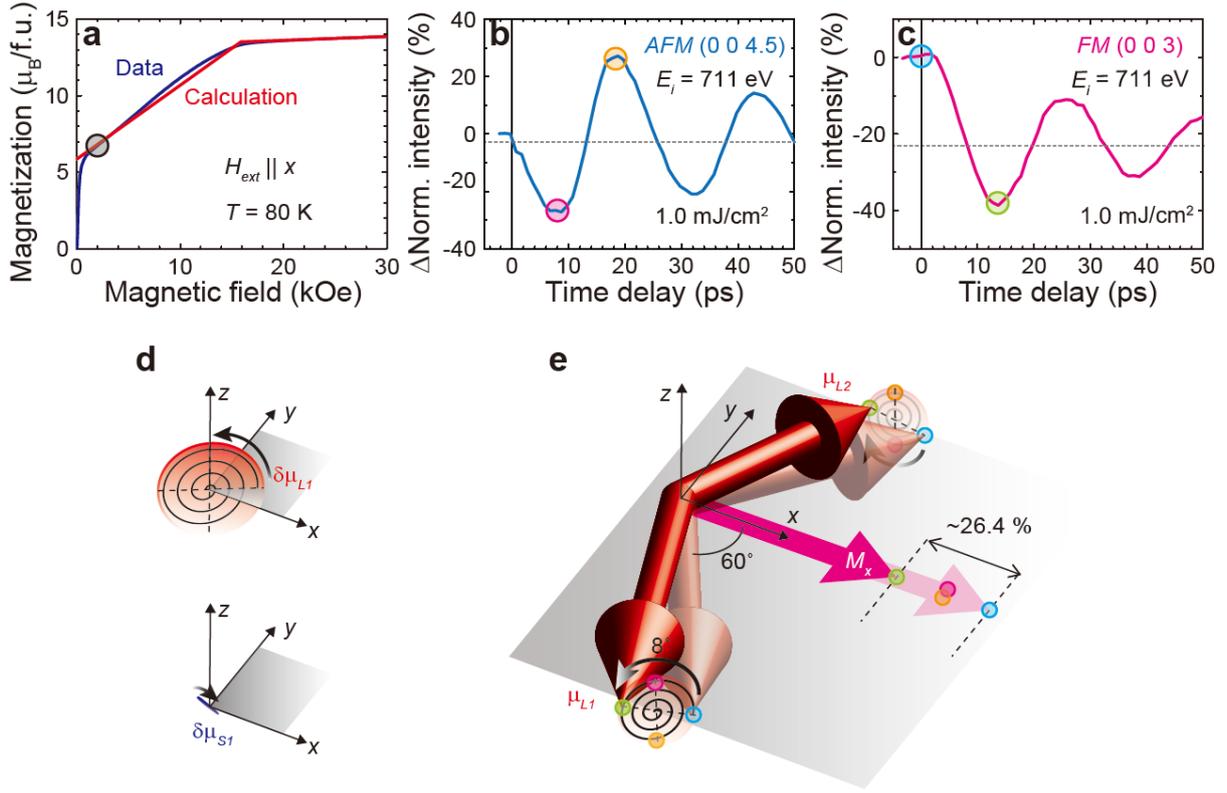

**Fig. 3| Net magnetization modified by the photoexcited *AFM* magnon.** (**a**) $H_{ext}$ dependence of the magnetization ($M_x$) per a formula unit (f.u.) at $T = 80$ K presenting the experimental data (blue) and calculation result (red) from the magnetic Hamiltonian in *Eq*. 1. The grey circle denotes $M_x$ at $H_{ext}$ (= 2kOe) where the X-ray diffraction experiment is performed. (**b**) *AFM* (0 0 4.5) and (**c**) *FM* (0 0 3) intensities at $H_{ext} \parallel +x$ and $T = 82$ K for the laser fluence = 1.0 mJ/cm² using incident X-ray energy, $E_i = 711$ eV. The colored circles correspond to the same colored $\mu_L$ directions on the precession trajectory shown in (**e**). (**d**) Schematic illustration of precession trajectories of $\mu_{L1}$ and $\mu_{S1}$ simulated from the magnetic Hamiltonian. (**e**) Illustration of the magnetization (dark pink arrow along the *x* axis) modified by the *AFM* magnon at the first half of the oscillation period.

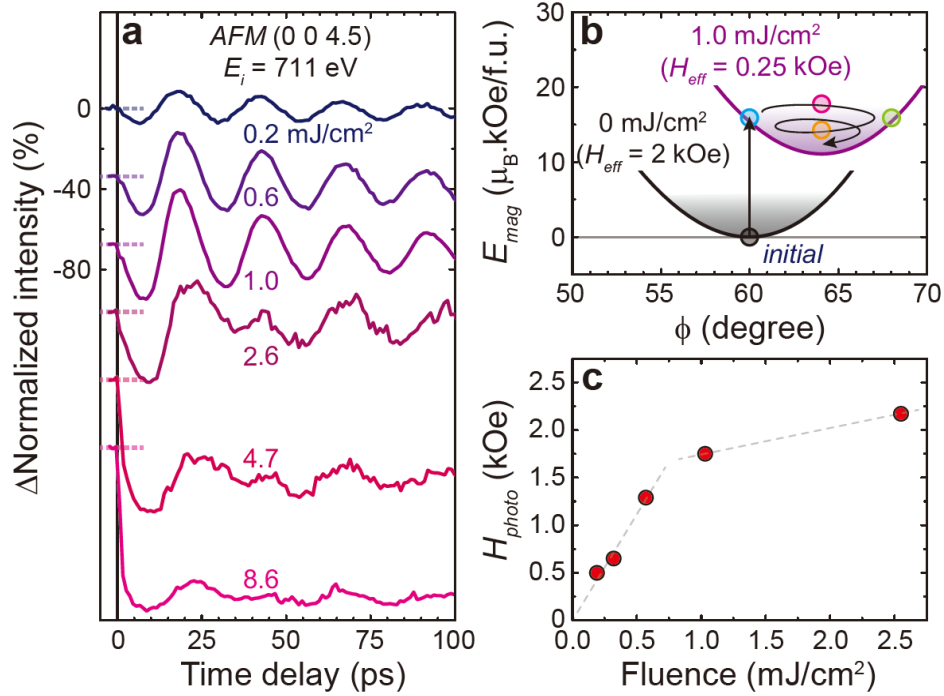

**Fig. 4| Effective magnetic-field changes due to laser excitation. (a)** *AFM* (0 0 4.5) intensities at $H_{ext}$ (= 2 kOe) ∥ +x and $T$ = 82 K using incident X-ray energy, $E_i$ = 711 eV. The data for each fluence are vertically shifted for clarity. The dashed lines denote initial intensities. **(b)** The magnetic potential energies calculated from *Eq*. 1 for $H_{eff} = H_{ext} - H_{photo}$ = 2 kOe (black) and 0.25 kOe (purple). The latter approximates the photoexcited energy for 1.0 mJ/cm². ϕ denotes the angle between $\mu_L$ and the x axis. The black arrow and spiral curve schematically illustrate the paths of the magnetic system being placed on high energy side of the photoexcited energy surface and moving towards the new energy minimum by generating the *AFM* magnon, respectively. The colored circles correspond to the same colored $\mu_L$ directions on the precession trajectory shown in Fig. 3. **(c)** The $H_{photo}$ ∥ -x for each fluence estimated from the coherent magnon trajectory. The dashed lines are guides to the eye.

**Methods**

**Sample preparation and characterization.** Single crystals of $Ba_{0.5}Sr_{1.5}Zn_2(Fe_{1-x}Al_x)_{12}O_{22}$ ($x = 0.08$) (BSZFAO) were grown from $Na_2O$-$Fe_2O_3$ flux in air. The chemicals were prepared with the molar ratio, $BaCO_3 : SrCO_3 : ZnO : Fe_2O_3 : Al_2O_3 : Na_2O = 19.69\ (1 - x') : 19.69\ x' : 19.69 : 53.61\ (1 - x) : 53.61\ x : 7.01$, heated up to 1420 ˚C, and underwent a series of thermal cycles. The $x'$ (= 0.85) and the thermal cycles followed the conditions in Ref. 47.

An epitaxial thin film of $BaFe_{10.2}Sc_{1.8}O_{19}$ (*M*-type hexaferrite) with thickness = 74 nm was grown on the $Al_2O_3$ (00.1) substrate by pulsed laser deposition under the same growth conditions as in Ref. 48 and annealed ex-situ twice, 2 hours at 930 ˚C and 2 hours at 1000 ˚C under flowing $O_2$ atmosphere. The film thickness and crystal structure have been confirmed by X-ray reflectometry and X-ray diffraction measurements (out-of-plane and Φ-scans), respectively, using a Panalytical X'Pert MRD diffractometer (Philips) equipped with a four bounce Ge (2 2 0) monochromator and a Cu $K_{\alpha 1}$ radiation source.

The magnetization vs. magnetic field curve for the BSZFAO single crystal was measured by using a VSM (vibrating sample magnetometer) incorporated with PPMS™ (Quantum Design).

**Time-resolved resonant magnetic X-ray diffraction experiment.** The optical laser pump and X-ray probe experiment was carried out at the SSS-RSXS endstation of the soft X-ray beamline of PAL-XFEL[49]. The BSZFAO single crystal was mounted on a 6-axis open-flow cryostat manipulator inside an ultra-high vacuum chamber. The sample was cooled down to $T = 82$ K with an external magnetic field ($H_{ext}$) = 2 kOe applied along the $x$ axis in the $ab$ plane. A pair of Nd-Fe-B magnets were attached on the sample holder and rotated together with the sample when the azimuthal angle was changed.

The optical laser pump was tuned to the wavelength = 400 nm, the pulse duration ~100 fs, and the repetition rate = 30 Hz. The laser was horizontally polarized (i.e., *p*-polarization) and was incident on the sample in nearly parallel to the X-ray probe. The time delay between the optical laser pump and the X-ray probe was controlled by a motorized linear stage.

The X-ray pulses were generated to have the pulse duration ~80 fs and the repetition rate = 60 Hz. The horizontally polarized X-ray and 2θ rotation of the detector established the horizontal scattering geometry resulting in incident π-polarization. The scattered X-ray pulses were detected by an APD (avalanche photodiode) whose signals were recorded by a high-speed digitizer. The incident X-ray photon energy was calibrated from the X-ray absorption spectrum (XAS) of 100-nm-thick Fe foil measured via the transmission mode across the Fe $L_3$ edge. The repetition rate of the X-ray twice that of the optical laser pump allowed us to compare the sample signals with/without the pumping for each time delay. This alternating probe removes errors from the mid- and long-term variations of the X-ray intensity.

**Penetration depth of the X-rays.** The X-ray penetration depth at $E_i$ = 702 eV, just below the Fe $L_3$ edge ~710 eV, is calculated to be 363 nm (density = 5.14 g/cm$^3$, incident X-ray angle = 90°) from the database of the Center for X-ray Optics[27]. It is well known that these calculations are accurate for X-ray energies away from the absorption edge. To estimate the penetration depth at the edge at 711 eV more precisely, we take into account the intensity ratio ~9.26 of the XAS spectrum[50] at the pre-edge and the absorption edge. The penetration depth is deduced to be 39 nm from dividing the depth at 702 eV by the XAS intensity ratio.

**Penetration depth of the optical laser.** Optical transmittance measurement by the UV-VIS-NIR (ultraviolet-visible-near infrared) spectroscopy identifies the absorption coefficient and penetration depth of the sample in the wavelength range. We attempted to measure the BSZFAO (*Y*-type hexaferrite) single crystal, but it was too thick (~200 μm) to obtain any relevant signal, absorbing almost all the incident photons. Alternatively, an epitaxial thin film of BaFe$_{10.2}$Sc$_{1.8}$O$_{19}$ (*M*-type hexaferrite) with thickness = 74 nm (Ref. 48) was measured to estimate the penetration depth. This *M*-type hexaferrite shares the crystal structure of the BSZFAO (*Y*-type hexaferrite) to a large extent and is known to have material properties in common such as the magnetic structure and magnetoelectricity[51]. Supplementary Fig. 1 shows a substantial rise of the absorption coefficient below the wavelength ~450 nm (blue, left axis), suggesting the electronic band gap. The penetration depth reduces abruptly around this wavelength and is counted to be 39 nm at the wavelength = 400 nm (red, right axis). It is expected that the BSZFAO has a similar magnitude of the penetration depth, while the onset of electronic absorption is ~600 nm (Ref. 35).

**Fitting of the transient magnetic Bragg peak intensities.** The magnetic Bragg peak intensities in Figs. 2a-c are fit with a damped oscillation and an exponential decay:

$$I = A_1 \exp(-\frac{t}{\tau_1}) \sin[2\pi(\frac{t-t_1}{\Omega})] - A_2 \exp(-\frac{t}{\tau_2}) + A_3$$

where $t$ is the time delay, $t_1$ is a phase shift, $\Omega$ is the oscillation period, and $\tau_1$ is a damping constant for the oscillation and $\tau_2$ the exponential decay constant; $A_1$, $A_2$ are amplitudes of the oscillation and the decay, respectively, and $A_3$ represent a constant background. The values extracted from the fit to the data in Figs. 2a-c are summarized in Supplementary Table 1.

**Magnetic Hamiltonian model calculation.** The static magnetization ($M_x$) vs. $H_{ext}$ relation is deduced from the magnetic Hamiltonian (*Eq.* 1) minimized with respect to the μ$_L$ and μ$_S$ directions[36]. The net moments of the *L* and *S* blocks are set as |μ$_L$| = 17.5 μ$_B$ and |μ$_S$| = 4 μ$_B$ with the notion that interlayer spins of Fe$^{3+}$ ions within each block are in antiparallel and unequal numbers of Fe$^{3+}$ sites at different layers result in the uncompensated moments. We find that the $M_x$ vs. $H_{ext}$ relation with $J_{LS}$|μ$_L$||μ$_S$| = 853 μ$_B$·kOe, $J_{LL}$|μ$_L$|$^2$ = 548 μ$_B$·kOe, $J_{SS}$|μ$_S$|$^2$ = (400 + 8 $D_S$) μ$_B$·kOe, and $H_{int}$ = 12.1 kOe is in good agreement with the experimental data (Fig. 3a).

The parameters and relation obtained above are employed to determine the magnetic anisotropy constants, $D_L$ and $D_S$. To this end, we use the Landau-Lifshitz-Gilbert equations (without damping terms for simplicity) that are written:

$$\frac{d}{dt}(\vec{\mu}_L) = -\gamma(\vec{\mu}_L \times \vec{H}_L) \text{ and } \frac{d}{dt}(\vec{\mu}_S) = -\gamma(\vec{\mu}_S \times \vec{H}_S)$$

where γ is the gyromagnetic ratio and $\vec{H}_{L(S)} = -\nabla_{\mu_{L(S)}} \mathcal{H}$ is the effective magnetic field on the $L(S)$ block. The solutions taking into account AFM precessions of $\mu_{L1}$ and $\mu_{L2}$, and $\mu_{S1}$ and $\mu_{S2}$ reveal two $k = 0$ normal AFM magnon modes[36]. We attribute the lower energy AFM magnon (low-lying mode) to the coherent magnon with the frequency of 41.7 ± 0.9 GHz. The higher energy AFM magnon (high-lying mode) is presumed in the frequency range 0.5 - 1.5 THz where a number of Y-type hexaferrites exhibit another AFM magnon mode[37,38]. By setting the frequency of this higher energy mode to 0.8 THz, we find $D_L|\mu_L|^2 = 116$ μ$_B$·kOe and $D_S|\mu_S|^2 = -544$ μ$_B$·kOe.

With all the parameters determined in the magnetic Hamiltonian, relative magnitudes of the AFM magnon modes are calculated. Supplementary Fig. 2 shows the trajectories of the lower- and higher-frequency modes. The former mode has dominant $\mu_L$ precession of which trajectory is nearly isotropic, while $\mu_S$ precession is insignificant and highly anisotropic along the $x$ axis. On the other hand, the latter mode exhibits comparable magnitudes of the $\mu_L$ and $\mu_S$ precessions that are largely elongated along the $z$ axis and nearly circular, respectively.

**Acknowledgements**

This work was supported by National Research Foundation of Korea (2019R1F1A1060295, 2019R1C1C1010034, 2019K1A3A7A09033399, 2019R1A2C2090648, 2020R1C1C1010477). Time-resolved resonant magnetic X-ray diffraction experiment were performed at the SSS-RSXS endstation of PAL-XFEL funded by the Korea government (MSIT). H.U. and U.S. were supported by the National Centers of Competence in Research in Molecular Ultrafast Science and Technology (NCCR MUST-No. 51NF40-183615) from the Swiss National Science Foundation and from the European Union's Horizon 2020 research and innovation programme under the Marie Skłodowska-Curie Grant Agreement No. 801459-FP-RESOMUS. P.B. was supported by the EPSRC under EP/H000925. M.J.R. thanks the Royal Society for the award of a Research Professor position.

**Author contributions**

H.J., H.-D.K., and S.H.C. performed the time-resolved resonant magnetic X-ray diffraction experiment. S.-Y.P. established the data aquisition system for the beamtime experiment. M.K., D.J., H.C., and I.E. prepared the optical laser pump setup. K.W.S. and K.H.K. grew the single crystals of the *Y*-type hexaferrite and characterized their magnetization. P.B. and M.J.R. grew and characterised the epitaxial thin fim of the *M*-type hexaferrite. H.J.S. measured the optical transmittance and reflectance of the samples. H.J. analyzed and simulated the oscillations in the magnetic Bragg peak intensities. S.H.C. performed the numerical calculations from the magnetic Hamiltonian. H.J., H.U., U.S., and S.H.C. interpreted the experiment and simulation results, and wrote the manuscript with contributions from all authors. H.J. and S.H.C. coordinated the project.

**Competing interests**

The authors declare no competing interests.

Supplementary Information

# 4D visualization of the photoexcited coherent magnon by an X-ray free electron laser

Hoyoung Jang[1,2], Hiroki Ueda[3], Hyeong-Do Kim[1], Minseok Kim[1], Kwang Woo Shin[4], Kee Hoon Kim[4], Sang-Youn Park[1], Hee Jun Shin[1], Pavel Borisov[5,6], Matthew J. Rosseinsky[6], Dogeun Jang[1], Hyeongi Choi[1], Intae Eom[1,2], Urs Staub[3] and Sae Hwan Chun[1,2]*

[1]*Pohang Accelerator Laboratory, Pohang, Gyeongbuk 37673, Republic of Korea*
[2]*Photon Science Center, POSTECH, Pohang, Gyeongbuk 37673, Republic of Korea*
[3]*Swiss Light Source, Paul Scherrer Institute, 5232 Villigen-PSI, Switzerland*
[4]*Center for Novel States of Complex Materials Research, Department of Physics and Astronomy, Seoul National University, Seoul 08826, Republic of Korea*
[5]*Department of Physics, Loughborough University, Loughborough, LE11 3TU, United Kingdom*
[6]*Department of Chemistry, University of Liverpool, Liverpool, L7 3NY, United Kingdom*
* Corresponding author: pokchun81@postech.ac.kr

**Contents**

**1. Transient intensity of (0 0 3) reflection at non-resonant photon energy**

**2. Time scale of the lattice thermalization**

**3. Laser fluence and temperature dependence of the *AFM* magnon precession**

**4. Magnetic Bragg peak intensities measured by resonant magnetic X-ray diffraction**

**5. Inspection of allowed coherent magnon precessions in the BSZAFO**

## 1. Transient intensity of (0 0 3) reflection at non-resonant photon energy

The resonant ferromagnetic (*FM*) (0 0 3) Bragg peak contains structure factors not only from the *FM* component of the canted antiferromagnetic (*AFM*) structure, but also from the crystal structure and orbital asphericity. If the oscillation in the transient (0 0 3) intensity would arise from a coherent phonon, we expect to see the oscillation from the crystal structure contribution at a non-resonant condition, where the magnetic contribution is absent. The incident X-ray photon energy was tuned to $E_i$ = 652 eV away from the X-ray absorption edges of the atomic elements constructing the BSZFAO. We observed no pronounced oscillation in the transient (0 0 3) Bragg peak intensity, but substantial intensity decrease after ~10 ps, which is saturated after ~50 ps (Supplementary Fig. 3). This intensity change shows a monotonic dependence with the fluence, indicative of photo-induced thermal effects with a combination of the Debye-Waller factor and the peak shift. This clearly supports our interpretation that the oscillations observed at the resonance are manifestation of a coherent magnon.

## 2. Time scale of the lattice thermalization

The *FM* (0 0 3) Bragg peak contains the crystal structure factor. Its profile of θ - 2θ scan serves as an indicator of the lattice change. Supplementary Figs. 4b & c show time-delay dependence of the peak shifts and widths obtained by fits to a Lorentzian for the θ - 2θ scans shown in Supplementary Fig. 4a. The Bragg peak shifts to a lower angle after ~10 ps, which manifests a lattice expansion. The width rapidly increases around the same time delay, indicating sizable strain gradient during the lattice expansion. The lattice expansion and the strain gradient imply that the lattice thermalization starts from the time delay ~10 ps.

## 3. Laser fluence and temperature dependence of the *AFM* magnon precession

The transient changes of normalized *FM* (0 0 3) Bragg intensities are presented in Supplementary Figs. 5a & b for various laser fluences. The intensities measured at $E_i$ = 702 eV display the nearly identical oscillation profile compared to those observed at resonance on top of the background that decreases with the fluence raised up to 8.0 mJ/cm$^2$. The oscillation frequency = 41.7 GHz is maintained as shown in the Fourier transformation of the transient intensities (Supplementary Fig. 5d). On the other hand, the intensities measured at $E_i$ = 711 eV exhibit an additional slower oscillation above 2.3 mJ/cm$^2$. This oscillation becomes progressively dominant with the fluence increased up to 7.4 mJ/cm$^2$. The presence of two oscillation components and their evolution are evident in their Fourier transformation (Supplementary Fig. 5e). Meanwhile, the overall intensities decrease substantially as the photoinduced increase of lattice temperature affects and melts the transverse conical magnetic order.

The oscillation profile of the transient *FM* (0 0 3) Bragg intensity was also checked while increasing the sample temperature (Supplementary Fig. 5c). As temperature increases, the crest of the oscillation starts to shift to a longer time delay and this phase shift evolves into a half cycle shift of a sinusoidal profile at 240 K. This shift is attributed to a new spin equilibrium position distinct from that at lower temperatures. In addition, we notice that the oscillation period gradually increases with the temperature elevated (Supplementary Fig. 5f).

**4. Magnetic Bragg peak intensities measured by resonant magnetic X-ray diffraction**

The incident X-ray in the horizontal scattering geometry has π-polarization while the scattered X-ray at magnetic Bragg reflections has both π'- and σ'-polarizations. The magnetic Bragg peak intensities are the sum of the individual intensities from the π-π' and π-σ' channels viewing *FM* (or *AFM*) moment, $\vec{M}$:

$$I = I_{\pi-\sigma'} + I_{\pi-\pi'},$$

$$I_{\pi-\sigma'} = B(\vec{k}_i \cdot \vec{M})^2,$$

$$I_{\pi-\pi'} = B(\vec{k}_f \times \vec{k}_i \cdot \vec{M} + C)^2,$$

where $\vec{k}_i$ and $\vec{k}_f$ are the incident and scattered X-ray wavevectors, respectively, $B$ is a constant of proportionality, and $C$ is isotropic charge-scattering/orbital-scattering amplitudes. The $C$ has finite values only for *FM* (0 0 3*n*) (*n*: integer) peaks.

The magnetic Hamiltonian model (*Eq.* 1 in the main text) elucidates the directions of $\mu_L$ and $\mu_S$ with canted angles □ and α from the *x* and *z* axes, respectively. It is determined that □ = 60° and α = 30° at $H_{ext}$ (∥ +x) = 2 kOe. We take into account this magnetic structure to simulate the oscillations in the magnetic Bragg peak intensities. The notion that the magnitude of $\mu_L$ is significantly larger than that of $\mu_S$ and contributes more to the magnetic structure factors for both *FM* and *AFM* Bragg reflections leads us to build an effective magnon precession model considering only $\mu_{L1}$ and $\mu_{L2}$. Before the photoexcitation, the $\mu_{L1}$ and $\mu_{L2}$ form the canted *AFM* structure with the *FM* (***m*** = **μ**$_{L1}$ + **μ**$_{L2}$) and *AFM* (***l*** = **μ**$_{L1}$ - **μ**$_{L2}$) components along the *x* and *y* axes in the *ab* plane, respectively as shown in Supplementary Fig. 6. The precession of **μ**$_L$ is projected on the ***m*** and ***l*** that modulate the magnetic Bragg peak intensities.

**5. Inspection of allowed coherent magnon precessions in the BSZAFO**

First of all, we examine a case that the photoexcitation induces periodic modulation of the $\mu_L$ magnitude rather than angular change (precession). In this case, both *FM* and *AFM* Bragg peak intensities are expected to modulate in the same phase, which turn out inconsistent with the observation in Fig. 2 (the main text). Hereafter, we inspect the magnon precessions allowed in the BSZFAO, including *FM* and *AFM* magnons, to find the one that accounts for the observed distinct oscillations in the *FM* and *AFM* intensities.

Supplementary Figs. 7-10 summarize the available precessions that display distinct oscillation and transient behaviours of the *FM* and *AFM* Bragg reflections. We consider the $\mu_{L1}$ and $\mu_{L2}$ precessions around quasi-equilibrium directions denoted as $\mu_{L1}'$ and $\mu_{L2}'$ with various conditions: (1) $\mu_{L1}'$ and $\mu_{L2}'$ moved out of the *xy* plane (= *ab* plane) or within the plane, (2) the same or opposite directional changes of $\mu_{L1}'$ and $\mu_{L2}'$, (3) clockwise or counter-clockwise $\mu_{L1}$ precession, and (4) *FM* or *AFM* magnon precessions of the $\mu_{L1}$ and $\mu_{L2}$. For the simulation, each angle between $\mu_{L1}$ and $\mu_{L1}'$, and between $\mu_{L2}$ and $\mu_{L2}'$ is set to be 3˚ with the precession period = 24 ps. Among these conditions, only the case illustrated in Supplementary Fig. 10a (also Fig. 2f in the main text) shows the behaviours consistent with the experimental results observed for the *FM* (0 0 3) and *AFM* (0 0 4.5) Bragg reflections. This situation with slight demagnetization (= 5 %) considered also matches the behaviour of the *AFM* (0 0 1.5) reflection.

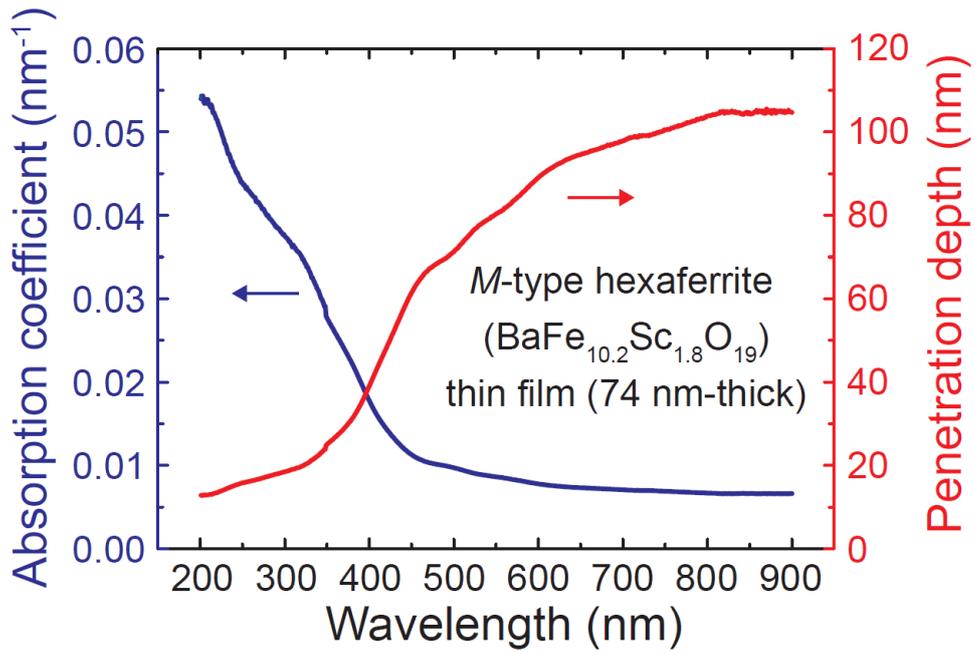

**Supplementary Fig. 1| Absorption coefficients and penetration depths of a *M*-type hexaferrite thin film (BaFe$_{10.2}$Sc$_{1.8}$O$_{19}$).** The epitaxial film with thickness = 74 nm was grown on the Al$_2$O$_3$ (00.1) substrate.

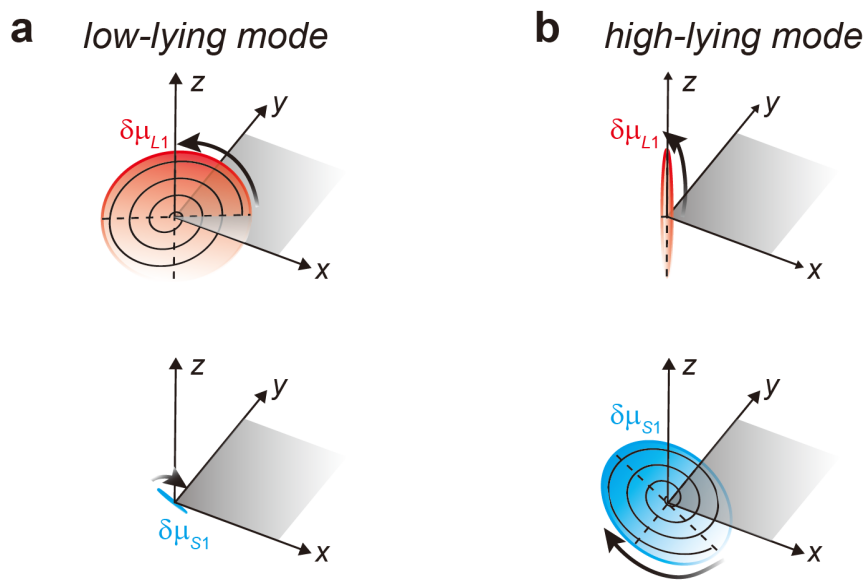

**Supplementary Fig. 2| Relative magnitudes of the *AFM* magnons calculated from the magnetic Hamiltonian.** (**a**) The lower energy magnon. (**b**) The higher energy magnon. The precession magnitude is normalized by the *z* axis-component of $\delta\mu_{L1}$. The curved black arrows denote the precession directions.

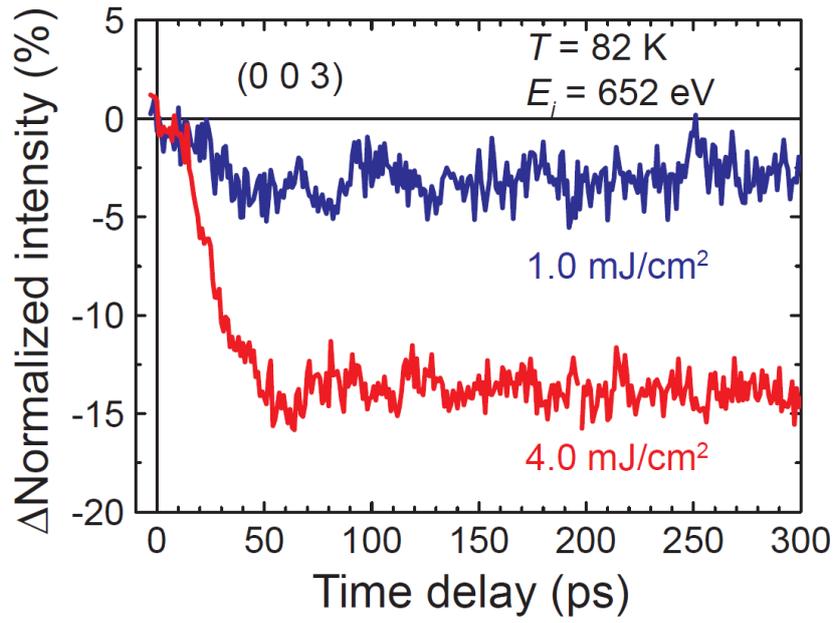

**Supplementary Fig. 3| Transient change of the normalized (0 0 3) Bragg peak intensity at a non-resonant condition, $E_i$ = 652 eV.** The data were obtained for the laser fluences = 1.0 (blue) and 4.0 mJ/cm² (red) at $T$ = 82 K under $H_{ext}$ = 2 kOe applied along the $x$ axis in the $ab$ plane.

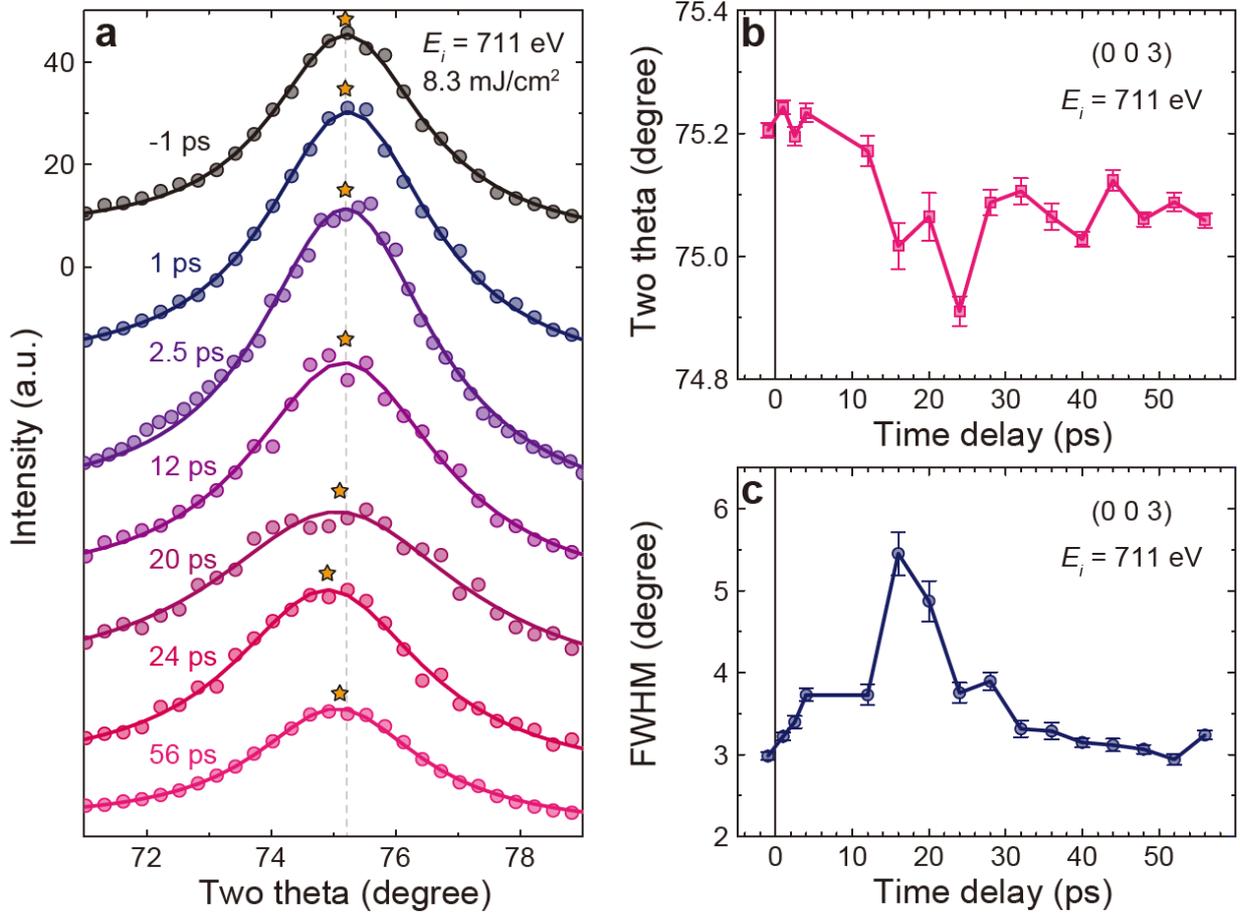

**Supplementary Fig. 4| θ - 2θ scans of the (0 0 3) reflection taken at $T$ = 82 K for high laser fluence = 8.3 mJ/cm$^2$ at selected time delays.** (**a**) The data (circle) fit with the Lorentzian profile (solid line) are vertically shifted for clarity. Each peak position is indicated by the star symbol. The vertical dashed line depicts the peak position at the time delay = -1 ps. Time dependence of the peak position (**b**) and peak width (**c**) in full-width-at-half-maximum obtained from the fits as a function of time delay.

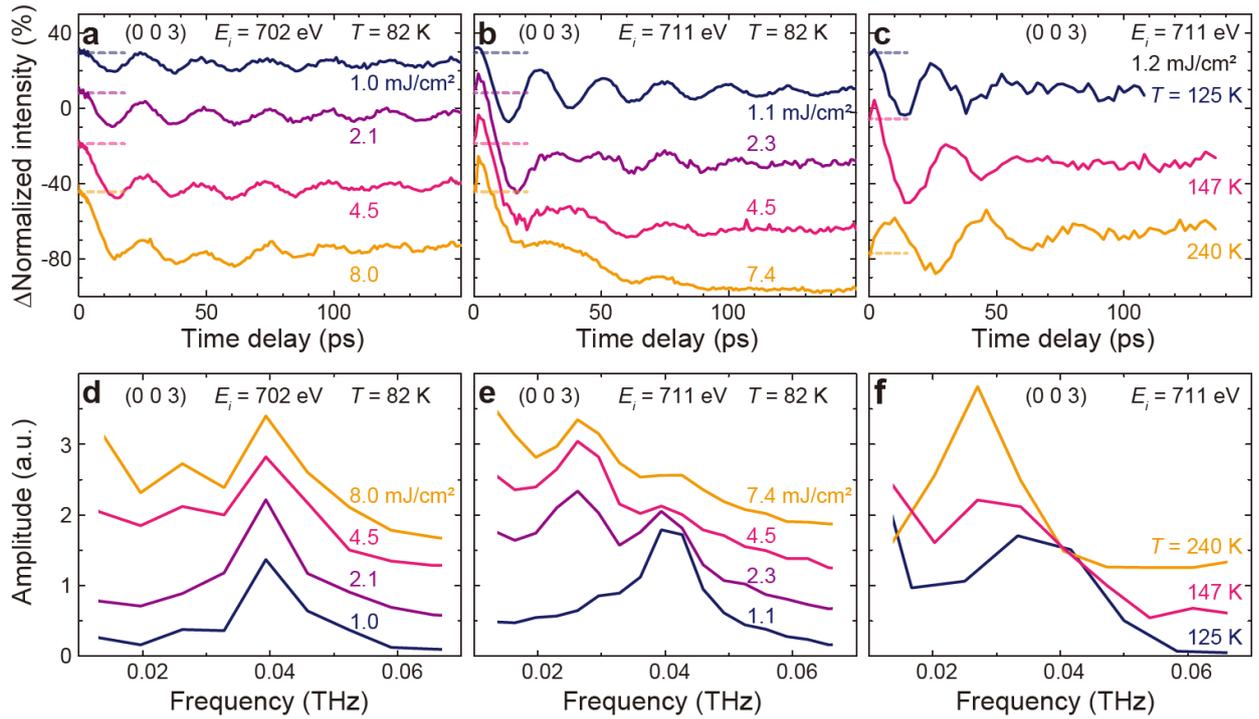

**Supplementary Fig. 5| Laser fluence and temperature dependences of transient change of the normalized *FM* (0 0 3) intensity.** Laser fluence dependence of the normalized intensity change at $T$ = 82 K for $E_i$ = 702 eV (**a**) and 711 eV (**b**). (**c**) Temperature dependence of the traces for 1.2 mJ/cm² and taken at $E_i$ = 711 eV. The data for each fluence and temperature are vertically shifted for clarity. The shifted base lines are indicated on the left side with dashed lines. (**d-f**) Fast Fourier transform of the intensities in (**a-c**). The transformed data are vertically shifted for clarity.

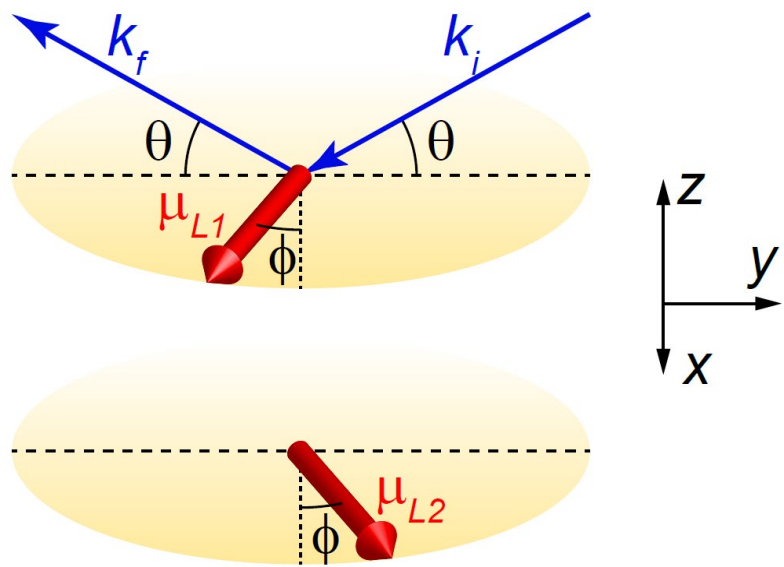

**Supplementary Fig. 6|** Schematic illustration of the canted *AFM* structure consisting of $\mu_{L1}$ and $\mu_{L2}$.

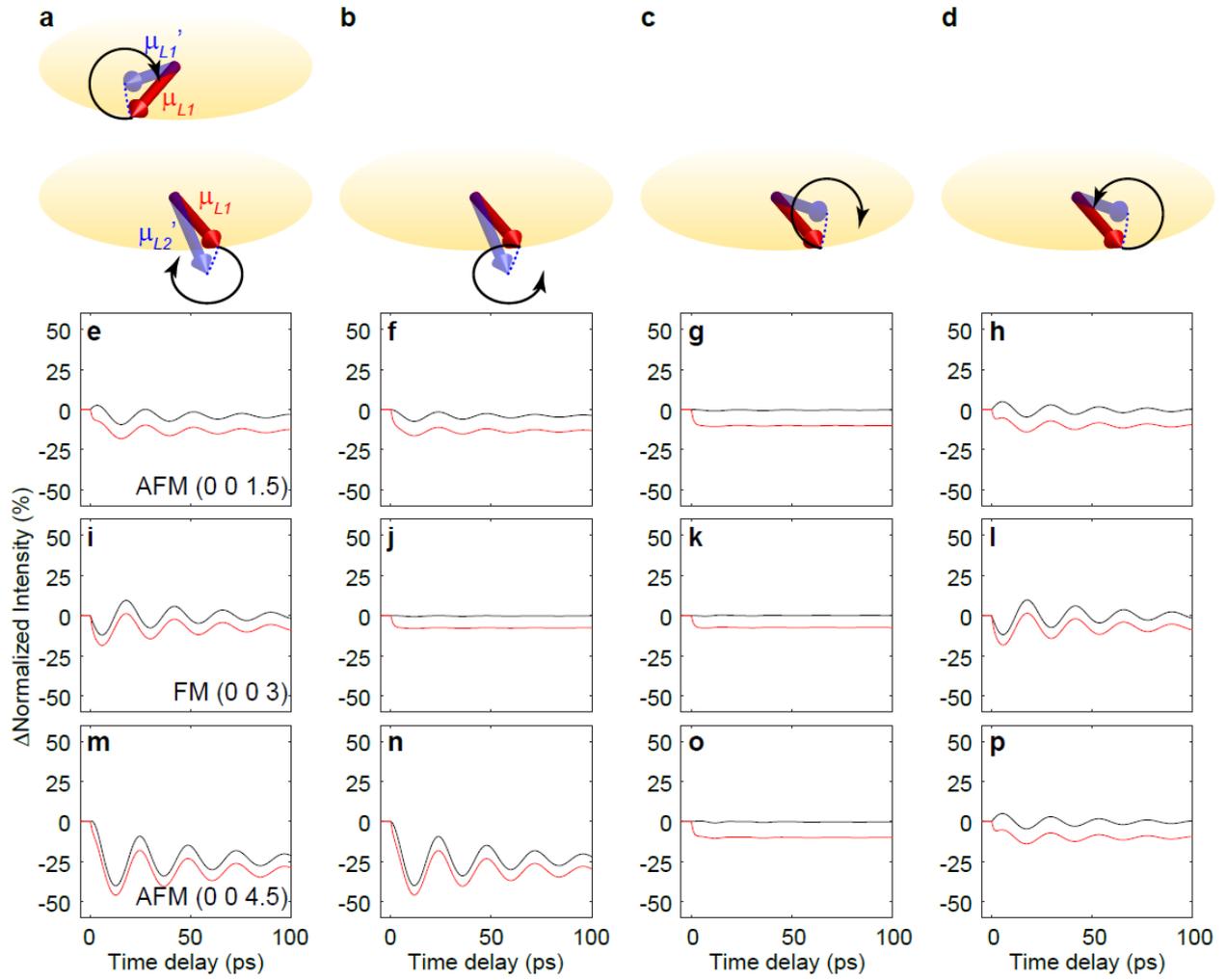

**Supplementary Fig. 7| Simulated transient behaviours of the *FM* and *AFM* Bragg reflections for the case where $\mu_{L1}$ precesses clockwise around a new quasi-equilibrium $\mu_{L1}'$ direction located out of the *xy* plane.** (**a**) Clockwise and (**b**) counter-clockwise $\mu_{L2}$ precessions around $\mu_{L2}'$ moved opposite to $\mu_{L1}'$ along the *z* axis. (**c**) Clockwise and (**d**) counter-clockwise $\mu_{L2}$ precessions around $\mu_{L2}'$ moved in the same direction with $\mu_{L1}'$ along the *z* axis. The red, blue, and black arrows denote $\mu_L$, $\mu_L'$, and precession directions, respectively. The transient intensities of (**e-h**) *AFM* (0 0 1.5), (**i-l**) *FM* (0 0 3), and (**m-p**) *AFM* (0 0 4.5) Bragg peaks for the cases illustrated in (**a-d**), respectively. The intensities are normalized with the one at a negative time delay. Black (red) curves present the simulation results without (with) initial demagnetization, respectively.

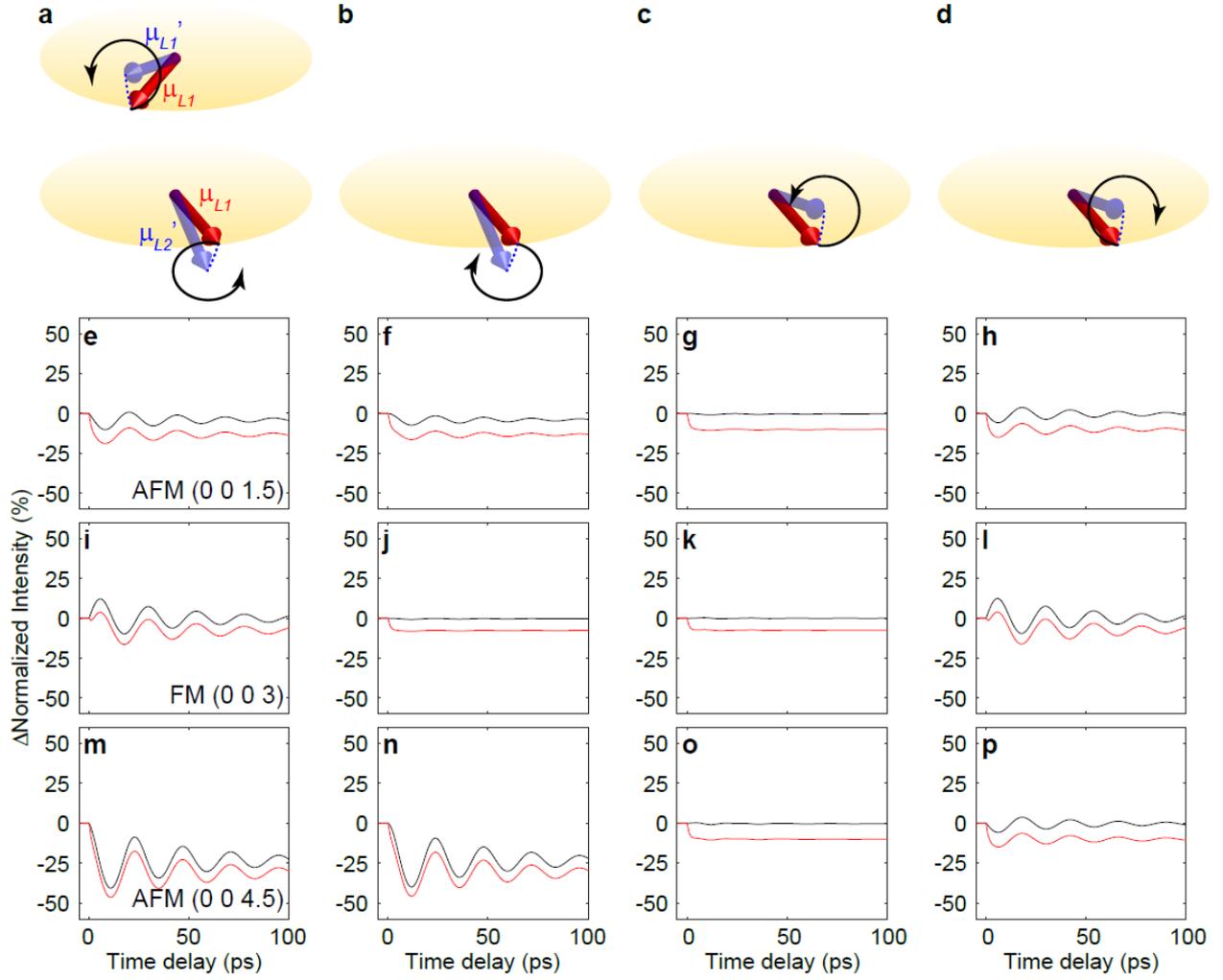

**Supplementary Fig. 8|** Simulated transient behaviours of the *FM* and *AFM* Bragg reflections for the case where $\mu_{L1}$ precesses counter-clockwise around a new quasi-equilibrium $\mu_{L1}'$ direction located out of the *xy* plane. (**a**) Counter-clockwise and (**b**) clockwise $\mu_{L2}$ precessions around $\mu_{L2}'$ moved opposite to $\mu_{L1}'$ along the *z* axis. (**c**) Counter-clockwise and (**d**) clockwise $\mu_{L2}$ precessions around $\mu_{L2}'$ moved in the same direction with $\mu_{L1}'$ along the *z* axis. The red, blue, and black arrows denote $\mu_L$, $\mu_L'$, and precession directions, respectively. The transient intensities of (**e-h**) *AFM* (0 0 1.5), (**i-l**) *FM* (0 0 3), and (**m-p**) *AFM* (0 0 4.5) Bragg peaks for the cases illustrated in (**a-d**), respectively. The intensities are normalized with the one at a negative time delay. Black (red) curves present the simulation results without (with) initial demagnetization, respectively.

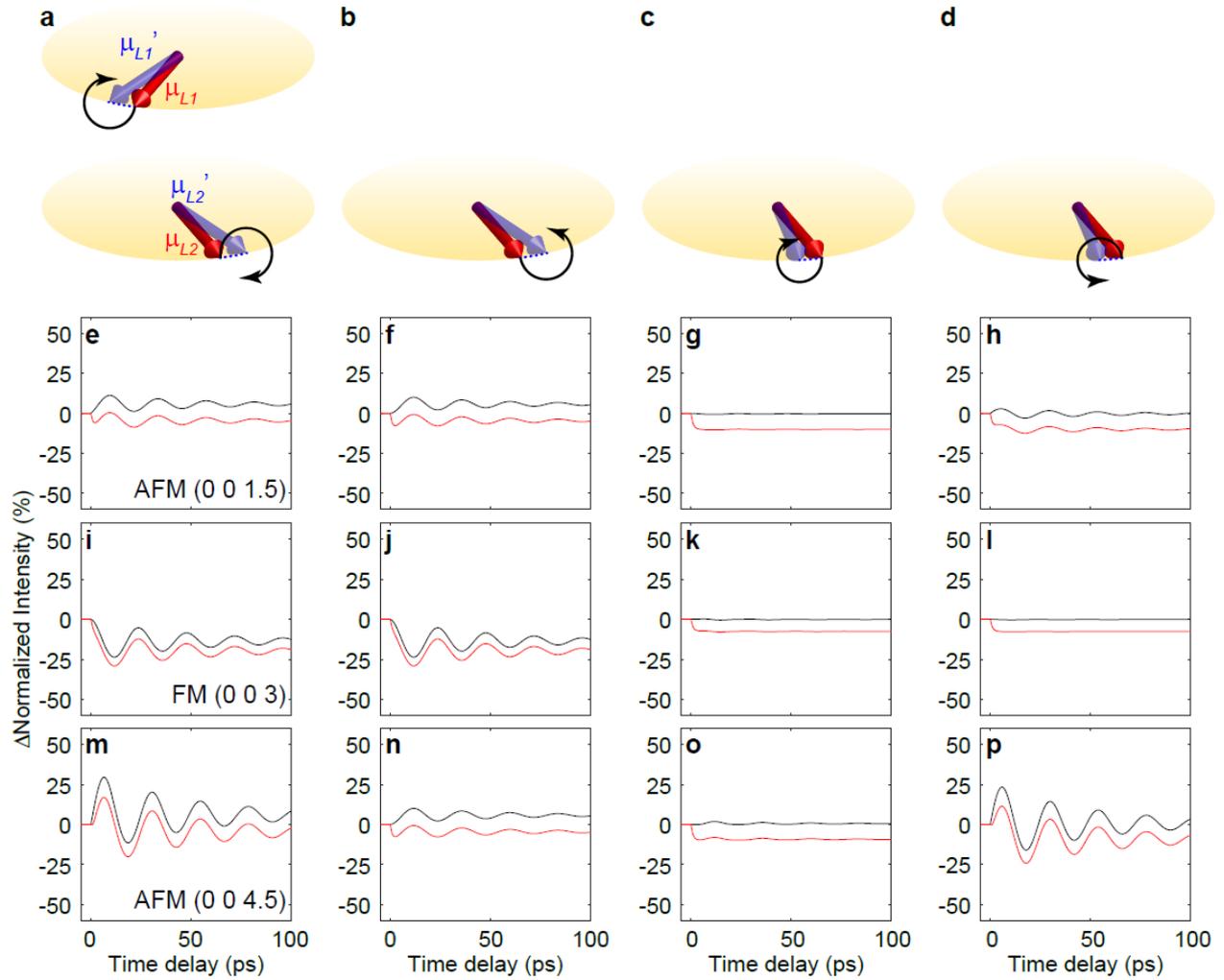

**Supplementary Fig. 9| Simulated transient behaviours of the *FM* and *AFM* Bragg reflections for the case where $\mu_{L1}$ precesses clockwise around a new quasi-equilibrium $\mu_{L1}'$ direction moved within the *xy* plane and tilted away from the *x* axis.** (**a**) Clockwise and (**b**) counter-clockwise $\mu_{L2}$ precessions around $\mu_{L2}'$ moved opposite to $\mu_{L1}'$ tilted away from the *x* axis. (**c**) Clockwise and (**d**) counter-clockwise $\mu_{L2}$ precessions around $\mu_{L2}'$ moved in the same direction with $\mu_{L1}'$ tilted away from the *x* axis. The red, blue, and black arrows denote $\mu_L$, $\mu_L'$, and precession directions, respectively. The transient intensities of (**e-h**) *AFM* (0 0 1.5), (**i-l**) *FM* (0 0 3), and (**m-p**) *AFM* (0 0 4.5) Bragg peaks for the cases illustrated in (**a-d**), respectively. The intensities are normalized with the one at a negative time delay. Black (red) curves present the simulation results without (with) initial demagnetization, respectively.

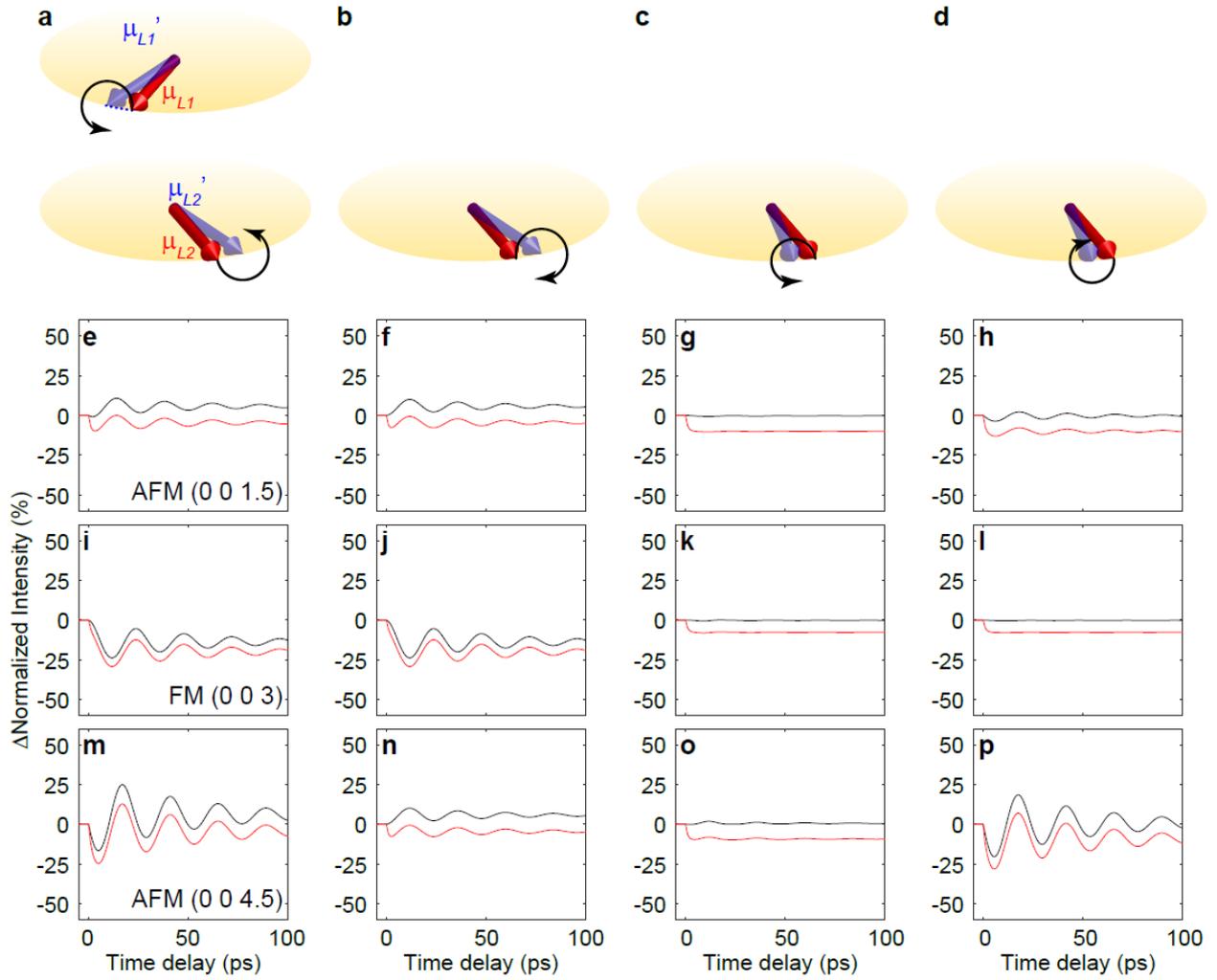

**Supplementary Fig. 10|** Simulated transient behaviours of the *FM* and *AFM* Bragg reflections for the case where $\mu_{L1}$ precesses counter-clockwise around a new quasi-equilibrium $\mu_{L1}'$ direction moved within the *xy* plane and tilted away from the *x* axis. (**a**) Counter-clockwise and (**b**) clockwise $\mu_{L2}$ precessions around $\mu_{L2}'$ moved opposite to $\mu_{L1}'$ tilted away from the *x* axis. (**c**) Counter-clockwise and (**d**) clockwise $\mu_{L2}$ precessions around $\mu_{L2}'$ moved in the same direction with $\mu_{L1}'$ tilted away from the *x* axis. The red, blue, and black arrows denote $\mu_L$, $\mu_L'$, and precession directions, respectively. The transient intensities of (**e-h**) *AFM* (0 0 1.5), (**i-l**) *FM* (0 0 3), and (**m-p**) *AFM* (0 0 4.5) Bragg peaks for the cases illustrated in (**a-d**), respectively. The intensities are normalized with the one at a negative time delay. Black (red) curves present the simulation results without (with) initial demagnetization, respectively.

|  | $A_1$ | $A_2$ | $A_3$ | $\tau_1$ (ps) | $\tau_2$ (ps) | $t_1$ (ps) | $\Omega$ (ps) |
|---|---|---|---|---|---|---|---|
| (0 0 1.5) | 0.06 | 0.07 | 1.00 | 119.77 | 2.81 | 12.18 | 23.96 |
| (0 0 3) | 0.06 | 0.02 | 0.94 | 106.47 | 22.09 | 18.85 | 24.07 |
| (0 0 4.5) | 0.10 | - | 0.99 | 107.39 | - | 12.38 | 24.24 |

**Supplementary Table 1| Fitting parameters for the transient Bragg peak intensities in Figs. 2a-c (the main text).** The $A_2$ for the *AFM* (0 0 4.5) Bragg peak is negligible making $\tau_2$ undefined.